\documentstyle[11pt]{article}
\include{epsf}

\begin{document}

\vbox{\vspace{20mm}}
\large
\begin{center}
{\bf Lattice Models for Magnetic Fluids: Correlations Between Order Parameters}
\bigskip
\end{center}

\normalsize
\begin{center}
Dorina ANDRU VANGHELI\footnote{Corresponding author. Fax: +40 56 196088.
E-mail: vangheli@quasar.uvt.ro.},
{Hora\c tiu COVLESCU}

Gheorghe ARDELEAN, Cristian STELIA\\
{\em {Department of Theor. and. Comp. Physics,\\ West University of Timi\c soara, V. P\^arvan Str., no. 4,
}}

{\em {1900 Timi\c soara, Romania}}\bigskip
\end{center}

\begin{abstract}
Magnetic fluids are colloidal suspensions of ferromagnetic particles covered with a surfactant layer, dispersed in a host liquid.
The existence of cooperative phenomena in such magnetic colloidal systems, makes
the determining of their physical properties a difficult task.
In the framework of statistical mechanics, taking into consideration interparticle interactions, 
the study of equilibrium properties and phase transitions of magnetic fluids can be best done. 

In this paper we analyze some basic aspects about the real possibility of 
describing the critical properties of these systems 
in the language of spin-1 lattice-gas model and we also propose the use of a higher spin (3/2) model. 
\end{abstract}

\hspace*{\parindent} The main features of magnetic fluids are related to their strong magnetic properties (magnetic
susceptibility, $\chi\sim$ 0.1 - 1).

For a long period of time, all theoretical approaches in the field relayed on hidrodynamical
models, thus regarding magnetic fluids as continuous media. The absence of interparticle 
interactions and the postulation of the Langevin law as a magnetic state equation, common to all
this models, let unexplained a number of phenomena.
Within these, of great practical and theoretical importance is the aggregation process, i.e.
the formation of large agglomerates of magnetic particles in the presence of a magnetic field. In most cases, these
agglomerates vanish when the field is switched off, which suggests that the process is reversible.
Meanwhile, deviations of magnetic properties (magnetization, initial susceptibility) from the Langevin behavior were 
observed as a consequence of aggregation.
The phenomena was recorded by various experimental methods (see for example \cite{take}).

A new approach, from the viewpoint of statistical mechanics, 
which takes into consideration the interparticle interactions, was needed.
Thereafter, statistical models appeared in the theoretical studies on 
magnetic fluids.

{\it Cebers}, \cite{cebers}, and, independently, {\it Sano} and {\it Doi}, \cite{sano}, emphasized that the occurrence of
agglomerates in the presence of an external magnetic field could be regarded as a 
phase separation generated by the thermodynamic instabilities of the colloidal solution. The description 
of these phenomena was given in the framework of the mean field theory.
Other approaches include the cell model \cite{calicu} and the hard dipolare spheres model \cite{zangu}.
These models led to remarkable results on the magnetic fluids equilibrium thermodynamic and on their phase transitions.

Another general approach was made by using the lattice-Boltzmann model on a triangular bidimensional lattice, \cite{sofo}. In fact,
te model is a particular case of the Ising-$\frac{1}{2}$ model. Is is well known, 
that for the Ising-$\frac{1}{2}$ triangular bidimensional structure, critical 
properties are determined by the fixed point Hamiltonian, it's dipolar coupling constant being $J=0.34$ in the absence of an external field.
The  density of particles for this value of the coupling constant is $n=10^{24}$ ferroparticles/m$^3$ \cite{noi}, a result which is in good agreement with the
critical value of the volume fraction. As a consequence of this observation, 
we can conclude that the Ising-$\frac{1}{2}$ model can be successfully used to analyze, qualitatively, both critical behavior and
stability properties of magnetic fluids. 

The development of theories on exact or approximate solutions of the Ising-$\frac{1}{2}$ models is based on the existence of a single order parameter,
identified with the density, concentration, crystalografic order parameter, magnetization or polarization of the system.

But, in the case of magnetic fluids, a good description of the agglomeration phenomena can not be completely specified 
by a single order parameter, because, in this case we have to deal 
with two types of ordering processes: magnetic and structural. 
In these conditions, we will need at least two order parameters, which can be done in using a Ising-$1$ lattice model (\cite{doi}, \cite{trei}).

If we consider the following general Ising-like Hamiltonian 
\begin{equation}
{\cal H}=-H\sum _i S_{i}-J\sum _{i,j}S_{i}S_{j}-D\sum _i (S_i)^2-K\sum _{i,j}(S_i)^2(S_j)^2
\end{equation}
where $Q_i=(S_i)^2$ is the singleparticle quadrupolar operator, as the two order
parameters we shall use the quantities
\begin{equation}
M=<S_i>
\end{equation}
and
\begin{equation}
Q=<(S_i)^2>
\end{equation}
These two parameters are not independent in the sense that, generally speaking,
$M\neq 0$ implies $Q\neq 0$.

In the Bragg-Williams mean-field approximation, for the spin-1 
general Hamiltonian, where $H=0$, the kinematical coupling between M and Q is determined, [3],
with the following expressions
\begin{equation}
M=\frac{2\exp \beta\left( D+2KQ \right) \sinh \left( 2\beta JM \right) }{1+2\exp \beta\left( D+2KQ \right) \cosh \left( 2\beta JM \right) }
\end{equation}
and
\begin{equation}
Q=\frac{2\exp \beta\left( D+2KQ \right) \cosh \left( 2\beta JM \right) }{1+2\exp \beta\left( D+2KQ \right) \cosh \left( 2\beta JM \right) }
\end{equation}
or
\begin{equation}
M=Q\tanh \left( 2\beta JM \right)
\end{equation}
where $\beta\equiv \frac{1}{kT}$.

It was observed, from the analysis of these expressions, that, as a function of
the $\frac{D}{K}$ ratio, there are possible dipolar, quadrupolar or successive transitions.
Assuming that $K$ favors condensed state and $J$ determines the ferromagnetic or
antiferromagnetic ordering, we consider that this type of analysis is important for the identification of
the microstructural properties which are responsible for the magnetic behavior of magnetic fluids.

For a more detailed analysis of the properties of such a system, we represent the Bragg-Williams free energy / site
in the form
\begin{eqnarray}
\Phi &=& -DQ - KQ^2 - JM^2 -kT \left\{ \ln{\cal N} -(1-Q)\ln[{\cal N}(1-Q)] -\right.\nonumber\\
&-&\left.\frac{1}{2}(Q+M)\ln [\frac{1}{2}{\cal N}(Q+M)]-\frac{1}{2}(Q-M)\ln [\frac{1}{2}{\cal N}(Q-M)]\right\}
\end{eqnarray}
where $\cal N$ represents the number of sites of the lattice, and, for its
equilibrium value, determined by the condition
\begin{equation}
\frac{\partial {\Phi}}{\partial Q}=0
\end{equation}
is obtained the condition
\begin{eqnarray}
&-&D - 2KQ - 2J\frac{1 - \frac{2K}{kT}Q(1-Q)}{\frac{2K}{kT}(1-Q)} - \frac{1}{kT}\cdot\nonumber\\
&&\left\{\ln 2(1-Q)-\ln\sqrt{(Q^2 - M^2)} - \frac{1}{2}\frac{1-\frac{2K}{kT}Q(1-Q)}{\frac{2K}{kT}M(1-Q)}\ln\frac{Q+M}{Q-M}\right\}
\end{eqnarray}
or
{\small
\begin{eqnarray}
&&(1+\mu)\nu D^2 + \frac{2\gamma\nu [\nu - 2Q(1-Q)]}{2(1-Q)}\cdot D +\nonumber\\
&&\mu\left[\ln 2 (1-Q) - \ln\sqrt{Q^2-M^2} - \frac{\nu - 2Q(1-Q)}{2M(1-Q)}\ln\sqrt{\frac{Q+M}{Q-M}}\right] = 0
\end{eqnarray}
}
where the following notations are used
\begin{equation}
\frac{2J}{kT} = \gamma~\mbox{,}~~~\frac{D}{K} = \frac{1}{\mu}~\mbox{,}~~~kT=\nu K.
\end{equation}

It was interesting for us to represent the curves $D=D(Q)$, for various values of T,
in different physical situations, determined by $\gamma$-values.

\begin{figure}[htbp]
\centerline{\epsfxsize11cm \epsfbox{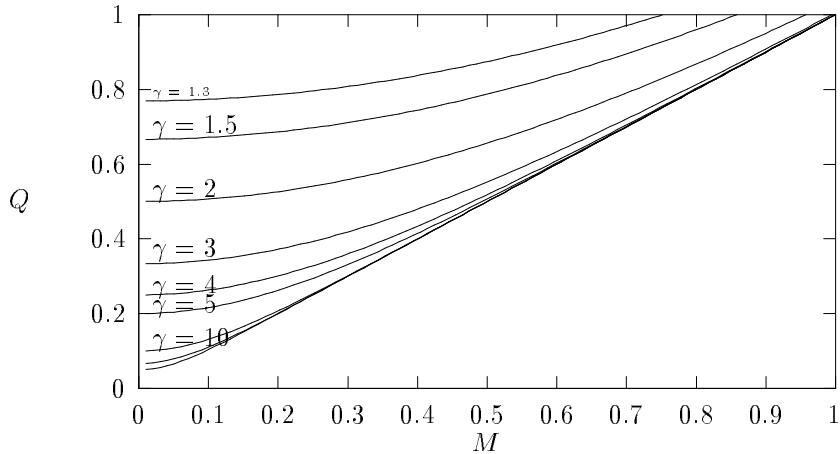}}
\caption{$Q=Q(M)$ dependence for various $\gamma$ values}\label{fig1}
\end{figure}

At the first stage, from figure \ref{fig1} we observe that, as a consequence of the mathematical (kinematical) relationship
between Q and M, the only relevant $\gamma$-values are in the range $1.1 < \gamma < 15$.
It is because, for $\gamma < 1.1$, the Q-order parameter values less than unity are not founded,
and, on the other hand, for $\gamma > 15$, the curves are $\gamma$- insensitive,
being almost of the same form. 

Our analysis reveals the $\gamma-$insensivity of the inflexion points for each 
$D=D(Q)$  which preserves $\mu$ in the domain $(-1,~0)$. 
Taking $\mu$ in the same domain, the analytical form of the curves
is almost insensible to variations of $\mu$.

Comparing the figures that show de dependency $D=D(Q)$, we can conclude that, for a
given temperature, the form of the curves (extremes and inflexion points) hence, the existence
of different types of phase transitions, is strongly influenced by the coupling constant $D$,
while the $J$-coupling constant determines the degree of order at which the transition occurs
(the increase of $Q$, and in the same time that of $M$).

It is obviously that the restriction for $H=0$ excludes the discussion about
the segregation phenomena in ferrofluids. The problem will be analyzed in a future paper. 

Considering the structural complexity of magnetic fluids we proposed, \cite{patru}, regarding
phase transitions, the use of an Ising-3/2 type model.
For that we assumed the existence of "free" ferroparticles, i.e. ferroparticles which, by means of various causes, are not completely covered with surfactant. 
The concentration of this "free" ferroparticles strongly influences the agglomeration process. 

In defining the model we assume the same cell construction, as used by Kalikmanov, \cite{calicu}, but with the needed specifications for a lattice model.
The above considerations imply the existence of two types of cells: cells within we have ferroparticles covered 
with a surfactant layer and cells which contain "free" ferroparticles.
Two spin states will be allocated for each type of cell: one state for the 
ferromagnetic particle with or without it's surfactant layer 
and the second state for the remaining solvent in the cell.
The main reason of this construction lies in the different 
dimensions of ferroparticles and solvent molecules (the first are much greater than the latter). 

In the general case, for the Ising-S model, the general form of the Hamiltonian is, \cite{cucu},
\begin{equation}\label{1}
{\cal H}=-H\sum_{i}S_i-J\sum_{ij}S_iS_j-J\sum_{ij}\left[ (S_i)^2 - \frac{1}{3}S(S+1)\right]\left[S_j-\frac{1}{3}S(S+1)\right]
\end{equation}

The Ising-3/2 Hamiltonian can now be obtained now considering $S=\frac{3}{2}$ in (\ref{1})
\begin{equation}\label{2}
{\cal H}=-J\sum_{ij}S_iS_j-K\sum_{ij}\left[ (S_i)^2 - \frac{5}{4}\right]\left[(S_j)^2-\frac{5}{4}\right]
\end{equation}

In the case $S=\frac{3}{2}$ it is also possible to 
introduce an "octupolar" ordering, with an order parameter given by $(S_i)^3$, but this
is not the case here, since the Hamiltonian contains no terms of the form $(S_i)^3(S_j)^3$.
If we introduce the molecular fields $H$ and $D$, associated with the two order parameters $M$  and $Q$, 
the one-particle Hamiltonian
in the mean field approximation will be
\begin{equation}
{\cal H}_0=-H\sum_{i}S_i-D\sum_{i}\left[ (S_i)^2 - \frac{5}{4}\right]
\end{equation}

As it was shown, the Ising-3/2 Hamiltonian 
can be rewritten so that, the equations for $M$ and $Q$ 
in the mean field approximation, became \cite{cucu}
\begin{eqnarray}
M&=&\frac{3e^{\beta (2K Q+D)}\sinh (3\beta JM + \frac{3}{2}\beta H) + e^{\beta (-2KQ+D)}\sinh (\beta J M +\frac{1}{2}\beta H)}
{2e^{\beta (2K Q+D)}\cosh (3\beta JM + \frac{3}{2}\beta H) + 2e^{\beta (-2KQ+D)}\cosh (\beta J M +\frac{1}{2}\beta H)}\\
Q&=&\frac{2e^{2\beta (2K Q+D)}\cosh (3\beta JM + \frac{3}{2}\beta H) - 2e^{\beta (-2KQ+D)}\cosh (\beta J M +\frac{1}{2}\beta H)}
{2e^{\beta (2K Q+D)}\cosh (3\beta JM + \frac{3}{2}\beta H) + 2e^{\beta (-2KQ+D)}\cosh (\beta J M +\frac{1}{2}\beta H)}
\end{eqnarray}

As a result, we get a greater number of situations in which phase transitions occur.

\begin{figure}[htbp]
\centerline{\epsfxsize12cm \epsfbox{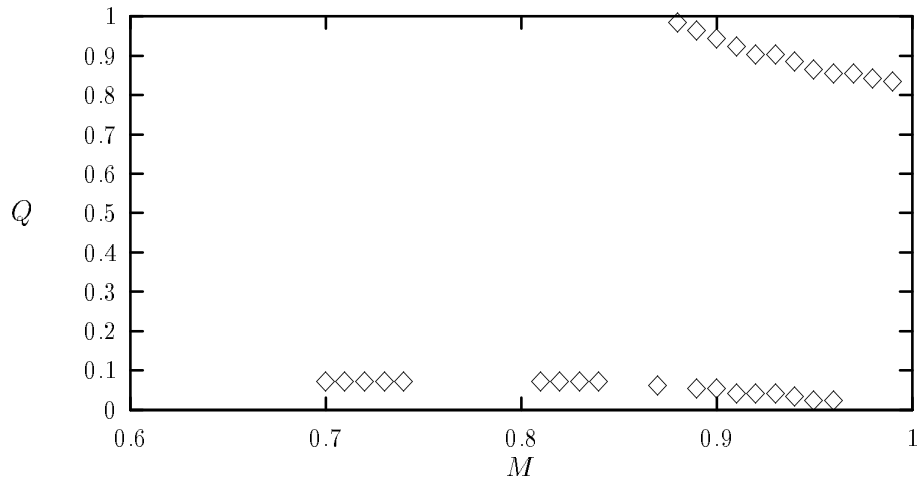}}
 \caption{$Q=Q(M)$ dependence}\label{fig2}
\end{figure}

Same considerations as in the spin-1 case, in the absence of an external magnetic field, led us to the conclusion that the $Q=Q(M)$ dependence is
strongly determined by the value of $D$, and, in some cases, we get two values for the order parameter $Q$, both corresponding to the same value of $M$.
As an example, for a given set of coupling constants in the domain of interest ($\gamma=1$, $\mu=0$), the $Q=Q(M)$ dependence has the form given in figure \ref{fig2},
from which we observe that, in addition to the $M$-discontinuity domain $0.74\div 0.8$, at $Q=0$, it also exists a nonanalitical behavior of $M$ in the domain
of configurational ordering with $Q\neq 0$.

{\bf {Acknowledgments.}} This work was carried out under grant no. 7012 with M.E.N.

\end{document}